\documentclass[5p,times]{elsarticle}

\journal{optics communications}

\usepackage{amsmath}            
\usepackage{epstopdf}           
\usepackage{flushend}           
\usepackage{hyperref}           
\usepackage{lineno,hyperref}
\usepackage{graphicx}
\usepackage{subfigure}
 \usepackage{color}
 \usepackage{stfloats}
\definecolor{Revise}{rgb}{0.00,0.00,0.00}
\definecolor{Revise2}{rgb}{0.00,0.00,0.00}
\usepackage{amssymb}

\biboptions{square,numbers,compress}
\usepackage[figuresright]{rotating}

\usepackage{subfigure}

\begin{document}
\newcommand{\tabincell}[2]{\begin{tabular}{@{}#1@{}}#2\end{tabular}}
\begin{frontmatter}

\title{Efficient 525 nm laser generation in single or double resonant cavity}
\author[mymainaddress,mysecondaryaddress]{Shilong Liu}
\author[mythirdaddress]{Zhenhai Han}
\author[mymainaddress,mysecondaryaddress]{Shikai Liu}
\author[mymainaddress,mysecondaryaddress]{Yinhai Li}
\author[mymainaddress,mysecondaryaddress]{Zhiyuan Zhou\corref{cor1}}
\cortext[cor1]{Zhiyuan Zhou is the corresponding author.}
\ead{zyzhouphy@ustc.edu.cn}

\author[mymainaddress,mysecondaryaddress]{Baosen Shi}
\address[mymainaddress]{Key Laboratory of Quantum Information, University of Science and Technology of China, Hefei, Anhui 230026, China}
\address[mysecondaryaddress]{Synergetic Innovation Center of Quantum Information \& Quantum Physics, University of Science and Technology of China, Hefei, Anhui 230026, China}
\address[mythirdaddress]{School of Physics and Mechanical \& Electrical Engineering, Hexi University, Zhangye,734000, China}

\begin{abstract}
This paper reports the results of a study into highly efficient sum frequency generation from 792 and 1556 nm wavelength light to 525 nm wavelength light using either a single or double resonant ring cavity based on a periodically poled potassium titanyl phosphate crystal (PPKTP). By optimizing the cavity$'$s parameters, the maximum power achieved for the resultant 525 nm laser was 263 and 373 mW for the single and double resonant cavity, respectively. The corresponding quantum conversion efficiencies were 8 and 77\% for converting 1556 nm photons to 525 nm photons with the single and double resonant cavity, respectively. The measured intra-cavity single pass conversion efficiency for both configurations was about 5\%. The performances of the sum frequency generation in these two configurations was studied and compared in detail. This work will provide guidelines for optimizing the generation of sum frequency generated laser light for a variety of configurations. The high conversion efficiency achieved in this work will help pave the way for frequency up-conversion of non-classical quantum states, such as the squeezed vacuum and single photon states. The proposed green laser source will be used in our future experiments, which includes a plan to generate two-color entangled photon pairs and achieve the frequency down-conversion of single photons carrying orbital angular momentum.
\end{abstract}
\end{frontmatter}
\section{Introduction}
Coherent green laser plays an important role in both scientific and technical fields, such as in nonlinear optics, atomic physics, spectroscopy, and atmospheric physics, as well as for applications in the medical field, for laser printing, and for displays \cite{lvovsky2009optical,superfine1990phase,alnis2000sum,polzik1992spectroscopy,tanimura2006generation,vollmer2014quantum,breigenbach1997measurement}. In the field of quantum information, stable and narrow-band continuous green lasers are widely used in parametric down conversion, two-color entangled photon pair generation, two-color continuous optical field excitation\cite{krapick2013efficient,guo2011generation}, or generation of  tunable infrared lasers in optical parametric oscillators. Green laser at 525 nm is particularly interesting as it can be used to generate two-color entangled photon pairs with wavelengths of 1550 and 795 nm, which is useful for quantum interfaces between fiber-based telecom-band photons and atom-based(\textsuperscript{87}Rb) quantum repeaters\cite{bussieres2014quantum}.
Some common methods to generate green lasers are via semiconductor laser diodes, second harmonic generation (SHG), and sum frequency generation (SFG)\cite{tawfieq2015efficient,muller2012efficient,guo2015singly}. Although high-power green laser diodes are easily obtainable, they suffer from some disadvantages such as multi-longitudinal modes and poor beam quality. SFG can overcome these difficulties and has been proven to be a very suitable technique to generate high-quality, narrow linewidth laser sources\cite{yue2009continuous,moosmuller1997sum,
dawson2006multi,janousek2005efficient,mimoun2008sum}. Usually there are three configurations for SFG: single pass configuration (SPC), in which both pump beams make a single pass through the nonlinear crystal\cite{nishikawa2009efficient,
johansson2004generation,tawfieq2015efficient}; single resonance configuration (SRC), \color{Revise}one of the pump beams is in resonance with a cavity\cite{bjarlin2013single,jensen2013generation,guo2015singly}; double resonance configuration (DRC), {\color{Revise}both pump beams are in resonance with the cavity} \cite{vance1998continuous,mimoun2008sum}. SPC is a simple technique, but the output power and overall conversion efficiency are very low. Using a PPLN waveguide can improve the single-pass conversion efficiency (SPCE); however, it is difficult for the waveguide to support the conversion of spatial modes or images. There are two types of cavity enhanced configurations, namely SRC and DRC. In SRC,{\color{Revise} only one pump beam is in resonance with the cavity}, while the other beam only makes a single pass through the cavity; this configuration can enhance the conversion efficiency in comparison with the SPC, but the enhancement factor is not very high. Meanwhile in the DRC, {\color{Revise}both pump beams are in resonance with the cavity}, leading to a higher conversion efficiency in comparison with the SPC and SRC for the same power level. For the SRC, only one pump beam is locked to the cavity and there are no stability problems; conversely, with the DRC both beams are in resonance with the cavity and the technique suffers from the stability problems that are usually associated with the locking technique,
\begin{table*}[htp]
  \centering
  \caption{Summary of continuous wave-SFG green laser light generation around 525 nm.}\label{1}
  \begin{tabular*}{\hsize}{|l|l|l|l|l|l|l|l|}
  \hline
  Configurations&Pump laser&\tabincell{l}{Pump power \\and wavelength}& \tabincell{l}{Output power \\and wavelength}&\tabincell{l}{Nonlinear\\conversion \\efficiency}&\tabincell{l}{Nonlinear \\crystal}&Year& \tabincell{l}{Reference}\\
  \hline
   SPC & \tabincell{l}{DFB:Er-doped\\ distributed-\\feedback fiber \\laser} &\tabincell{l}{15+7.7W\\1565+783nm}& 1.2W,522nm & $1.0\%/W$ & \tabincell{l}{30mmPPSLT\\(HC Photonics)}& 2011 & \cite{vasilyev2011compact} \\
   \hline
    SPC & \tabincell{l}{DBR-tapered\\ diode lasers} &\tabincell{l}{ 7.8+7.8W\\1063+1062nm}& 3.9W,531nm & 2.6\%/W &\tabincell{l}{30mmPPMgLN\\(HC Photonics)} & 2012 & \cite{muller2012efficient}\\
    \hline
    SPC & \tabincell{l}{Two tapered diode\\ lasers} & \tabincell{l}{6.17+8.06W\\978+1063nm} & 1.7W,509nm & 4.3\%/W&\tabincell{l}{20mmPPMgLN\\(Covesion)} & 2015 & \cite{tawfieq2015efficient}\\
  \hline
    SRC & \tabincell{l}{ECDL: external-\\cavity diode laser} & \tabincell{l}{1.5+6.8W\\780+1560nm} & 0.268W,520nm & 1.2\%/W & \tabincell{l}{25mmPPKTP\\(HC Photonics)} & 2015 & \cite{guo2015singly} \\
    \hline
     SRC& \tabincell{l}{Ti: sapphire;\\diode laser}& \tabincell{l}{0.586+1.09W\\792+1556nm} & 0.263W,525nm & 5\%/W & \tabincell{l}{19mmPPKTP\\(Raicol)}& 2017 & \tabincell{l}{This\\ work} \\
     \hline
     DRC& \tabincell{l}{Ti: sapphire;\\diode laser}& \tabincell{l}{0.530+0.162W\\792+1556nm} & 0.373W,525nm & 5\%/W & \tabincell{l}{19mmPPKTP\\(Raicol)}& 2017 & \tabincell{l}{This\\work} \\
     \hline
  \end{tabular*}
\end{table*}
i.e. one needs to reprocess the error signals to achieve stable locking of both lasers\cite{mimoun2008sum}. Another difference between the SRC and DRC is the bandwidth of the output laser; for the DRC, the bandwidth of the output laser is dependent on the cavity$'$s bandwidth and thus limited to a few MHz. By comparison, the bandwidth in the SRC can reach several gigahertz, depending on the bandwidth of the input infrared laser. Both the SRC and DRC are suitable for high-efficiency SFG for a variety of applications. {\color{Revise2}{Table 1 presents the details of several recent reports on the progress in generating continuous green laser light using SFG, including the achieved output powers and nonlinear conversion efficiencies as well as listing which nonlinear crystals were used. With regard to DRC-SFG, a near 400 mW, 589 nm laser was constructed in 1998 \cite{vance1998continuous}; in 2008, SFG of 589 nm light was demonstrated with an efficiency close to unity\cite{mimoun2008sum}. However, for the generation of green laser light at 525 nm the use of such a configuration has been not yet reported to the best of our knowledge.}}

In this paper, we report on the results of a systematic study on cavity enhanced SFG with both the SRC and DRC. We propose detailed theoretical models for optimizing the cavity parameters in both configurations. In our experiments, we studied the performances of cavity enhanced SFG in both configurations using the optimized cavity parameters. The experimental results are in good agreement with our theoretical model's predictions. With the DRC, 373 mW, 525 nm (green) laser light was obtained when the power of the coupled input signals were 530 and 162 mW for the 792 nm (master) and 1556 nm (infrared) light, respectively; the corresponding intra-cavity quantum conversion efficiency (QCE) was 77\%. For the SRC, the achieved maximum green laser power was 263 mW when the coupled pump powers were 586 mW (master) and 1.094 W (infrared); the corresponding QCE was 8\%. In both configurations, the intra-cavity SPCE was up to $5\%/W$, which is (to the best of our knowledge) the highest single-pass conversion efficiency for green laser light generation based on a PPKTP crystal (see Table 1). These results will provide guidelines for designing cavity enhanced SFG for quantum frequency conversion of single photon states or squeezed vacuum states.

\section{Determination of optimized cavity parameters}
SFG-based second-order nonlinearity requires three mixing waves to interact in a nonlinear crystal. In this process, the energy $(\omega_3-\omega_1=\omega_2)$ and momentum conservation conditions $(k_3-k_1-k_2-2\pi/\Lambda=0)$ must be satisfied. Usually, there are three configurations for SFG, which are shown in Fig. 1; they are the SPC, SRC, and DRC. In the following subsections, we will describe the theoretical model for these configurations.
\subsection{SPC}
 SPC is the simplest setup for generating green laser light (see Fig. 1 (a)). When the interacting waves are Gaussian beams, the resulting green laser's power can be expressed as follows: \cite{boyd1968parametric,guha1980effects}.
      \begin{equation}
        P_3=\frac{4\omega_1\omega_2\omega_3d_{eff}^2l}{\pi\epsilon_0c^4n_3^2}
        h(\sigma,\xi)e^{-\beta l }P_1P_2
     \end{equation}
 where $\omega_i(i=1,2,3)$ are the angular frequencies of three mixing waves; $k_i(i=1,2,3)$ are the wave vectors, where the labels 1, 2, and 3 represent the master, infrared, and green laser for simplicity. $c$ is the speed of light in vacuum; $\epsilon_0$ is the permittivity of vacuum; and $\beta=\sum{\beta_i/2}(i=1,2,3)$ is the total absorption coefficient per unit distance inside the crystal. Further, $l$, $n_i(i=1,2,3)$, $\Lambda$, and $d_{eff}$ are the length, refractive index, grating pole period, and effective nonlinear coefficient of the nonlinear crystal, respectively. $h(\sigma,\xi)$ is the Boyd and Kleinman focusing parameter\cite{boyd1968parametric}, which includes the spatial phase mismatching factor $\sigma(=\Delta kb/2)$ and focusing parameter $\xi(=l/b)$, which can be expressed as:
         \begin{equation}
          h(\sigma,\xi)=\frac{1}{4\xi}\iint\frac{e^{i\sigma(\tau-\tau')}}{(1+i\tau)(1-i\tau')}d\tau d\tau'
         \end{equation}
where $b={2\pi\omega_1^2n_1}/{\lambda_1}$ is the confocal parameter of the master laser. $h(\sigma,\xi)$ varies with the focusing parameter $\xi$ and is within $10\%$ of the maximum value for $1.56<\xi<5.31$.
     \begin{figure}[htp]
          \centering
          \includegraphics[width=8.5cm,height=7.2cm]{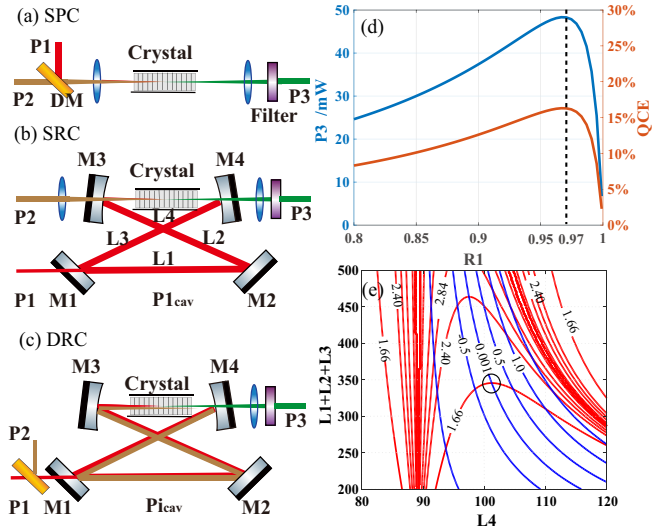}
          \caption{Three configurations for SFG and the optimized parameters for the SRC. Panels (a), (b), and (c) show the setups for the SPC, SRC, and DRC, respectively. In those panels, the following labels were used. DM: dichromatic mirror; filter: band filter; M1, M2: plane cavity mirrors 1, 2; M3, M4: concave mirrors 3, 4 (radius of curvature = 80 mm); crystal: PPKTP; and L1--L4: the distance of between the various mirrors. In panel (d), P3 (blue line) and QCE (red line) are plotted as a function of R1 in the SRC setup. Panel (e) shows the focal parameters (red line) and stable values (blue line) with regard to different bow-tie cavities geometric parameters for the master laser.}
         \end{figure}
\subsection{SRC}
    Based on Eq. 1, the output P3 can be simply written as $P_3=\alpha\cdot P_1 P_2$. $\alpha$ is the single pass conversion efficiency (SPCE) . In the SPC, the output P3 is weak because the SPCE is small\cite{nishikawa2009efficient}. To achieve a high power for the SFG, we need to increase the infrared or master laser's power.
    Here we employed a single resonant ring cavity to improve the circulating power of the master laser $P_1^{cav}$; a diagram of the infrared laser's single-pass through the crystal is shown in Fig. 1(b). The circulating power $P_1^{cav}$ is equal to the coupled input power $P_1$ multiplied by the enhancement factor $\Gamma_1$. $\Gamma_1$ depends on the reflectivity $R_1$ of the input mirror M1 and the round-trip intensity attenuation factor $r_1^2(r_1^2<1)$. Neglecting the absorption of the crystal, $\Gamma_1$ can be expressed as:
         \begin{equation}
         \Gamma_1=(1-R_1)/\left(1-\sqrt{r_1^2(1-C_1)}\right)^2
         \end{equation}
   where $C_1(=\lambda_3 P_3/ \lambda_1 P_1^{cav})$ is the coupling strength. The overall intensity attenuation factor $r_1^2$ can be roughly estimated by the {\color{Revise}{finesse}} of the cavity\cite{saleh1991fundamentals}, which can also be written as $(1-T_1)(1-\delta_1)$, where $T_1$ is the transmittance loss of M1 and $\delta_1$ is the overall loss in the cavity (ignoring $T_1$) for the master laser. The infrared laser make a single pass through the crystal and the circulating power $P_2^{cav}$ can be written as \cite{guo2015singly}:
           \begin{equation}
                P_2^{cav}=(1-\delta_2)(1-C_2)\cdot P_2
           \end{equation}
Here, $\delta_2$ and $C_2$ have the same expression for the infrared laser as for the master laser. From Eqs. 3 and 4 we can obtain the quantum conversion efficiency (QCE) $\eta=P_3^{fact}/(\lambda_2P_2/\lambda_3)$. {\color{Revise}{For a fixed value of $\delta_1$}}, there exists an optimum reflectivity $R_1$ that maximizes the QCE in this configuration. {\color{Revise2}{Considering that Eqs. 3 and 4 are complex coupling equations associated with the input powers, the first step is to study the relationship between the optimum value of $R_1$ and the input powers. We constructed a 2D color map using a linear interpolation (Fig. 2(c)) of the optimum value of $R_1$ versus P1 and P2, where $\delta_1=0.03$, $\delta_2=0.02$, and $\alpha=0.04/W$. The sampling interval of P1 and P2 was 10, and the corresponding sampling interval of $R_1$ for a group (P1, P2) was 50. The units of P1 and P2 are milliwatts [mW]. The color bar on the right represents the scale of $R_1$. In Fig. 2(c) it can be seen that the optimum values of $R_1$ changes slightly when the input power is varied between 100 and 1000 mW for the SRC.

\begin{figure}[htp]
          \centering
          \includegraphics[width=8.5cm]{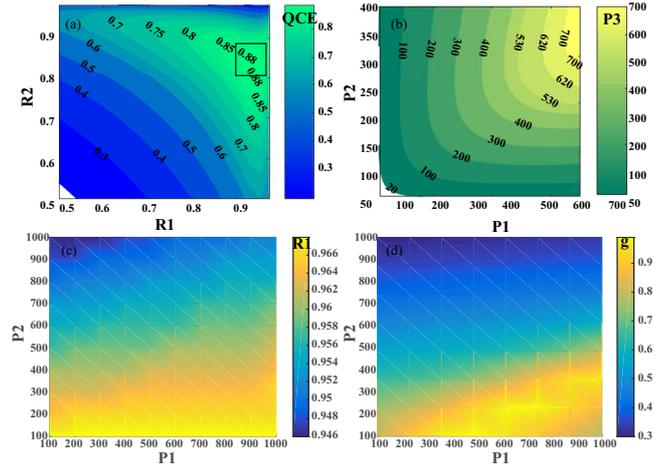}
          \caption{Contour map of the QCE and P3 for the SRC and DRC. {\color{Revise}{The simulation parameters were $\delta_1=0.03$, $\delta_2=0.02$, and $\alpha=0.04/W$ for the results presented in Fig. (a)-(d).}} Panel (a) shows the QCE versus R1 and R2. Panel (b) shows a plot of P3 versus P1 and P2, {\color{Revise}{where $R_1$=93\% and $R_2$=85\%.}} {\color{Revise2}
          {Panels (c) and (d) show the optimized reflectivity of the input mirror versus the input powers for the SRC and DRC, respectively.}}}
         \end{figure}

Next we studied the output power P3 and QCE versus $R_1$ for a specific set of input powers, namely P1 = 500 mW and P2 = 100 mW; }}Fig. 1(d) shows this relationship, where $\delta_1=0.03$, $\delta_2=0.02$, and $\alpha=0.04/W$. P3 and QCE will reach their maximum values when $R_1=97\%$. In fact, the optimized reflection values demonstrated that the impedance matching condition is given for the ring cavity\cite{torabi2003efficient,le200575}. In other words, the input power coupled to the cavity can attain its maximum.
In addition to finding the optimum value of $R_1$, we also needed to consider the beam's waist size and the stable conditions for the ring cavity. The various geometric parameters (L1, L2, L3, and L4) of the cavity determine the different focusing parameters ($\xi=l/b$) and the cavity's stable conditions ($|A+D|$). The stable conditions were calculated based on the ABCD matrix method\cite{saleh1991fundamentals}. To find the optimized parameters, we plotted the data shown in Fig. 1(e), where the red and blue lines represent the focusing parameters and the stable conditions, respectively; the horizontal axis is the distance of the two curved mirrors (L4) and the vertical axis is the remaining distance (L1 + L2 + L3). In the calculations, the parameters for the wavelength, the crystal length, and the curvature of the concave mirrors were 792nm, 19, and 80 mm, respectively. We chose the point where $|A+D|=0$ and $\xi=1.66 (h(\xi)=0.9558)$; the point is marked by a circle in Fig .1 (e). Under this condition, the beam's waist size inside the center of the crystal was 38 $\mu m$.
\subsection{DRC}
  To achieve high values of P3 or of the quantum conversion efficiency (QCE), a double resonant cavity is required. Fig. 1(c) illustrates this setup. P1 and P2 are coupled into the cavity by a dichroic mirror (DM). When {\color{Revise}the cavity is simultaneously in resonance with the two lasers} both circulating powers $P_i^{cav}(i=1,2)$ are enhanced, and the enhancement factor $\Gamma_i$ has a similar expression as in Eq. (3):
     \begin{equation}
         \Gamma_i=(1-R_i)/\left(1-\sqrt{r_i^2(1-C_i)}\right)^2
         \end{equation}
 where $C_i(P_i^{cav})=\lambda_3 P_3/ \lambda_i P_i^{cav}$. The optimization of the parameters is more complex for the DRC than for SRC because of the double resonance. {\color{Revise2} {As the same of SRC, we first needed to study the impact of the input powers on the optimum input reflectivity values $R_1$ and $R_2$. To simplify the task we defined a parameter $g(=P_3^{aver}/P_3^{max},0<g<1)$ to represent this complex coupling relationship, where $P_3^{max}$ is the maximum green laser output power under a given group of input powers, and $P_3^{aver}$ is the average output power for a given range of reflectivities, where the range of $R_1$ and $R_2$ were selected to be 0.92-0.96 and 0.82-0.88 ,respectively. The results are shown in Fig. 2(d), where $\delta_1=0.03$, $\delta_2=0.02$, and $\alpha=0.04/W$; the sample interval for P1 and P2 was 10; and the corresponding sample interval with $R_1 (0.7-0.99)$ and $R_2 (0.7-0.99)$ for a group (P1, P2) was 20 owing to limitations in the personal computer's calculation capacity. From Fig. 2(d), we can determine that it is impossible for us to realize a highly efficient DRC-SFG covering the entire range of input powers for a specific group of reflectivities. However, in most cases the infrared laser is weak, even on a single photon level\cite{albota2004efficient}. Next, we studied the QCE versus $R_1$ and $R_2$ while fixing one group of input powers (P1 = 500 mW, P2 = 100 mW).}} The dependence of QCE and P3 on $R_1$, $R_2$, and P1, P2 were simulated and the results presented in Fig. 2(a) and (b), respectively. In Fig. 2(a) it can be seen that the QCE reaches a maximum value (near 80\%) when the optimum parameters from the indicated square area in Fig. 2(a) are chosen, where $R_1$ is $92\%-96\%$ and $R_2$ is $82\%-88\%$, respectively. Based on the simulated QCE in Fig. 2(a), we selected $R_1=93\%$ and $R_2=85\%$ and plotted a contour map for the DRC (Fig. 2(b)), where the x(y)-axis is the input power P1(2) and the color depth represents the green laser's output power. The other simulation parameters for Fig. (a)-(d) are {\color{Revise}{$\delta_1=0.03$, $\delta_2=0.02$, and $\alpha=0.04/W$}}.
 \begin{figure*}[!htbp]
  \centering
  \includegraphics[width=16cm]{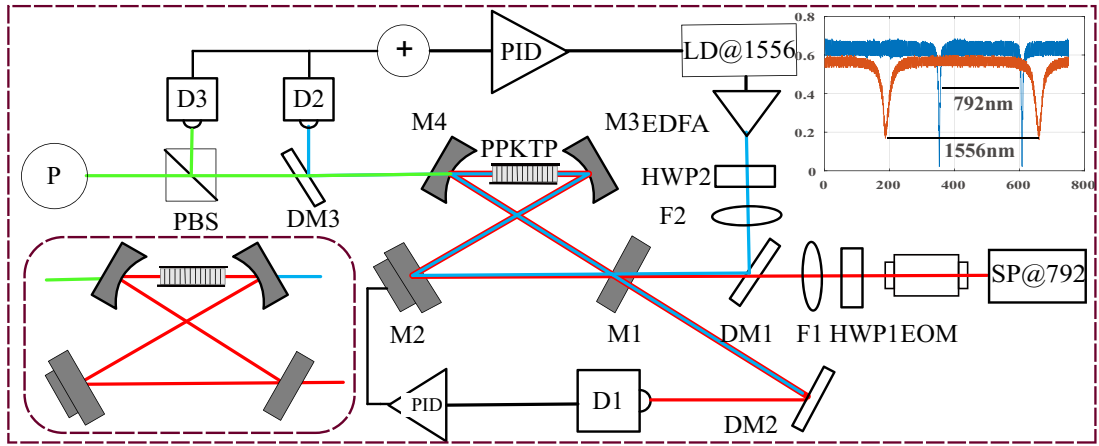}
  \caption{The experimental setup for SFG. The following labels were used in the figure. EOM: electro-optical modulator; HWP1 and HWP2: half wave plates 1 and 2; DM1-3: dichromatic mirrors 1-3; D1, D2, and D4: fast photodiodes1, 2, and 4; PBS: polarizing beam splitter; PID: proportional-integral-derivative controller; P: power meter (PM100D S142C, Thorlabs); and PPKTP: periodical pole potassium titanyl phosphate (supplied by Raycol Crystals, period of $9.375\mu m$). The lower left-hand corner of the diagram shows the setup for the SRC. In the top right-hand corner the reflection spectra for the 792 and 1556 nm lasers for the DRC are shown.}\label{a}
\end{figure*}
\section{{\color{Revise}{Experimental setups}} for the SRC and DRC}
The experimental setups for the SRC and DRC are shown in Fig. 3. The master laser is a Ti:sapphire laser (MBR110, Coherent). The infrared laser is a diode laser (Toptica pro design, wavelength tuning range of 1520-1590 nm) seeded onto an erbium-doped fiber amplifier (C-band, 1530-1560 nm). The PPKTP used was a 19-mm-long crystal with a cross section of $1\times 2mm^2$, which was located at the waist position of two curved mirrors; further, both end faces had anti-reflective coatings for the three wavelengths used in the study.

In the DRC, the cavity {\color{Revise}is simultaneously in resonance with two fundamental lasers.} The locking scheme is shown in Fig. 3. The first step is to lock the length of the cavity to the master laser. An electro optical modulator is driven by a radio frequency modulated local oscillator (3.15 MHz) to create sidebands of the master laser. The reflected light from M1 and DM2 is detected by a fast photon detector D1 and mixed with the split radio frequency modulated local oscillator to generate an error signal. The error signal is fed into a homemade PID circuit to lock the cavity to the master laser using PDH methods\cite{drever1983laser}. The next step is to lock the infrared laser to the cavity. A linear combination of the transmission spectrum of P2 and P3 is used to overcome a pronounced dip when the cavity is in double resonance\cite{mimoun2008sum}. The transmission spectrum is detected by D2 and D3 with the modulated current, which is fed into the infrared laser controller to lock the infrared laser onto the cavity. The reflectivities of the four mirrors were designed using optimization theory for the DRC, where $R_1=93\%$ and $R_2=85\%$ for M1 and M2-M4 had a high reflectivity over 99.9\% at two wavelengths of master and infrared lasers. M3 and M4 are two curved mirrors with the same curvature of 80 mm. M4 has a high transmission for the output green laser.

    For the SRC, the ring cavity (lower left-hand corner in the dashed box) {\color{Revise} is only in resonance with} the master laser (red line) and the infrared laser (blue line) via a single-pass through the crystal from M3. In this configuration, $R_1=97\%$ for P1 based on Fig. 1(d). Two curved mirrors, M3 and M4, have high reflectivity coatings at the master laser wavelength and a coating with a high transmission for the infrared and green lasers.
\section{Sum frequency generation}
The {\color{Revise}{finesse}} and coupling coefficient of the cavity can be estimated by measuring the reflection spectrum from D1. We show the two reflection spectra for the DRC in the top right-hand corner of Fig. 3. The blue and red spectra are from the master and infrared laser, respectively; here, the infrared spectrum was acquired when the cavity was locked to the master laser. From the reflectivity spectrum we can estimate the input coupling efficiency for P1 and P2. {\color{Revise}{For the DRC, the coupling efficiency for the two lasers was around 85\%; meanwhile, for the SRC, the infrared laser (P2) had a single pass through the crystal and the coupling efficiency was 100\%, while it was 85\% for P1. The corresponding reflection spectrum of P1 was similar to that of P1 for the DRC}}. The output power P3 was measured using a power meter P (PM100D S142C, Thorlabs). To estimate the intra-cavity power of the green laser, linear attenuation factors including the dichroic mirror (DM) and band pass filter were taken into account. The experimental and theoretical results for the SRC and DRC are shown and compared in this section.

  For SRC-SFG, the measured {\color{Revise}{finesse}} of the cavity for the master laser was 90, and therefore the intra-cavity depletion $\delta_1$ was equal to 0.039. The infrared laser did a single pass of the crystal; $\delta_2$ was around 0.02. The circulating power $P_1^{cav}$ in the cavity can be estimated by measuring the leaked power from M2 when the cavity is locked. Fig. 4(a) shows that the intra-cavity green laser output power (linear attenuation factor of 0.93) and QCE changed along with the coupled master laser's power when the infrared laser power was fixed at 1094 mW. The left y-axis of Fig. 4(a) shows P3 (blue stars) and the right y-axis shows QCE (red asterisks). The uncertainty of the measurement was around 5\% because of the floating input power and the measurement errors. The blue and red lines show the theoretical results for P3 and QCE based on Eqs. (3) and (4), respectively. {\color{Revise}{The corresponding plots of P3 versus the intra-cavity power ($P_1^{cav}$ and $P_2^{cav}$) are presented in Fig. 4(b), where the x-, y-, and z-axis shows $P_1^{cav}$, $P_2^{cav}$, and $P3$, respectively.
  }}

To observe the high power generated via SRC-SFG, a 3D solid line histogram was plotted (Fig. 5(a)). The histogram illustrates the output power in the experiment as a function of the two fundamental input powers, where the x-, y-, and z-axis are P1, P2, and P3, respectively. For SRC-SFG, {\color{Revise}{263 mW green laser light}} is generated when the input powers for the 792 and 1556 nm lasers were 586 and 1094 mW, respectively. The 3D dashed line histograms in Fig. 5(a) represent the theoretical predictions. {\color{Revise}{The parameters in the simulation were $\delta_1$=0.039, $\delta_2$=0.02, and $\alpha=0.05$.}}

 For DRC-SFG, {\color{Revise}{the cavity is simultaneously resonant with two fundamental laser wavelengths}}. The {\color{Revise}{finesse}} of both beams was 60 and 35, respectively (see top right in Fig. 3), corresponding to intra-cavity depletion values of $\delta_1=0.0347$ and $\delta_2=0.0295$, respectively. In this setup, the value of P3 and the QCE were measured using the power meter (P) when the cavity was simultaneously in resonance with both fundamental lasers. {\color{Revise}{The linear attenuation factor for the green laser was around 0.78, which includes the losses from DM3, the band pass filter, the polarizing beam splitter, and the collimating lens.}}

  Figs. 4(c), 4(d), and 5(c) show the results for the DRC. In Fig. 4(c), the infrared laser was fixed at a power of 161 mW and the measured power P3 (left y-axis) and QCE (right y-axis) are shown as a function of the increase in power of the master laser. The corresponding solid line shows the theoretical results calculated based on Eq. (5), {\color{Revise2}{where the intra-cavity depletion values were $\delta_1=0.0347$, $\delta_2=0.0295$, and $\alpha=0.05$, respectively}}. In those figures, we can find that a QCE of over 80\% was achieved when the power of the master laser was as high as 400 mW. {\color{Revise}{Fig. 4(d) shows the graph of P3 versus the intra-cavity powers for a fixed value of P2 of 161 mW.}} As for the SRC in Fig. 5(a), 3D solid and dashed line histograms are shown in Fig. 5(c) to illustrate the dependence of P3 on P1 and P2 in the experimental studies and the theoretical calculations. In this figure the maximum values of P3 and of the QCE reached 373 mW and 80\%, respectively. From Figs. 4(c) and 5(c) we can see that the experimental results are in good agreement with the predictions reported on in section 2, which demonstrates that our optimal cavity parameters are effective for SFG. But for SRC in Figs. 4(a) and 5(a), there exist a dislocation between theoretical and experimental results, which is because the reflection of input coupled mirror is not the optimized value under the high power level of infrared laser (see Fig. 2(c)).

    Compared with SRC-SFG, it is easy for DRC-SFG to achieve high power green lasing and high QCE when the input powers are the same level. {\color{Revise2}{For both the SRC and DRC the conversion efficiency shows the same increasing trend in Figs. 4(a) and 4(c): in the low pump power regime, the conversion efficiency has a good linear response with pump power P1, while the conversion efficiency tends to saturate in the high pump power regime. The differences between the two configurations lies in the maximum pump power needed to reach the maximum QCE. With the SRC, one laser (P1) is in resonance with the cavity, while the other laser (P2) makes a single pass through the crystal. The theoretical model for this configuration is given by Eqs. 3 and 4. In this configuration, in the low input power regime, P3 and QCE have a relatively good linear dependence on P1. If the input power is steadily increased, the conversion efficiency increases gradually. In fact, for the SRC, the maximum power for a QCE of unity has been theoretically calculated in \cite{zhou2016orbital}; however, for the DRC, the infrared laser must also be in resonance with the cavity, and the intra-cavity power of the infrared laser has to be amplified. Thus, the value of P3 and of the QCE for the DRC is higher than for the SRC with the same input power level. In the high input power region the values for P3 and for the QCE are saturated, mainly because most of  infrared photons were converted to green photons. In the saturation regimes, the values of P3 and QCE versus P1 have a slowly tendency. When we continued to increase the input power, the theoretical predictions show only a slight decrease in the P3 values and of QCEs; this is because the values of R1 and R2 were not the optimized values for the high power level used (see Fig 2(d)).}}
    \begin{figure} [tp]
          \centering
     \includegraphics[width=9.0cm]{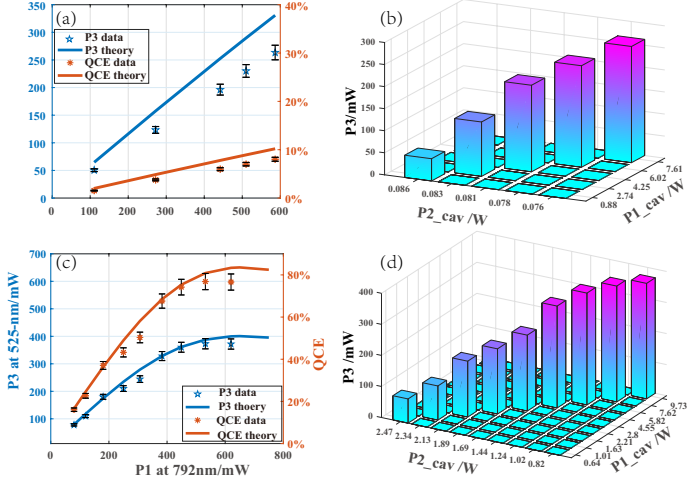}
          \caption{The results for SRC-SFG and DRC-SFG. (a) P3 and QCE were varied by changing P1 (P2 was fixed at 90 mW) for the SRC, {\color{Revise2}{where $\delta_1$=0.039; $\delta_2$= 0.02; and $\alpha=0.05$.}} (b) P3 versus $P_1^{cav}$ and $P_2^{cav}$ (P2 was fixed at 90 mW). (c): P3 and QCE were varied with P1 (P2 was fixed at 161 mW) for the DRC, where $\delta_1=0.0347$, $\delta_2=0.0295$, and $\alpha=0.05$. (d) The corresponding plot for P3 versus $P_1^{cav}$ and $P_2^{cav}$ for the DRC (P2 was fixed at 161 mW).
          }\label{1}
        \end{figure}
     {\color{Revise}{In our experiment, we also found that the locking system had a poor robustness under high input powers as high powers caused thermal effects to change the state of the cavity, including changing the absorption losses, focusing parameters, and phase matching conditions.}} Although the QCE was low with the SRC setup, an excellent robustness and a stable green laser emission were easily realized. By decreasing the intra-cavity overall passive losses, optimizing the focusing parameters, increasing the mode-coupling, and selecting a highly nonlinear crystal the QCE and output powers could be further improved with both the SRC and DRC.
 \begin{figure} [tb]
          \centering
     \includegraphics[width=9cm]{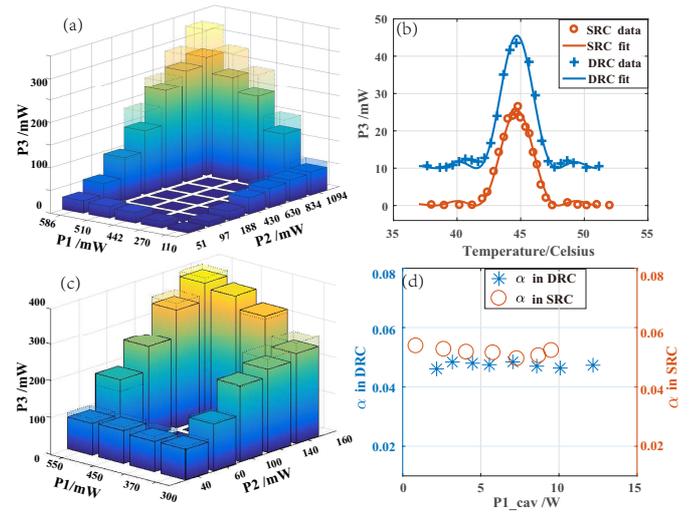}
          \caption{
          3D histograms for the SRC and DRC and the characteristic of the PPKTP crystal. Panels (a) and (c) show 3D solid and dashed line histograms to illustrate the dependence of P3 on P1 and P2 in the experimental and theoretical calculation for the SRC (with $\delta_1$=0.039, $\delta_2$= 0.02, and $\alpha=0.05$) and for the DRC ($\delta_1$=0.0347, $\delta_2$=0.0295, and $\alpha=0.05$). (b) The relationship between the tuned temperature and P3 is shown for the SRC (red line) and DRC (blue line). (d) The single-pass conversion efficiency ($\alpha$) of the PPKTP crystal for the SRC and DRC.
          }\label{1}
        \end{figure}

    The temperature of the PPKTP crystal was controlled by a homemade semiconductor Peltier temperature cooler with a stability of $\pm 2mK$. Fig. 5(b) shows the relationship of P3 and the tuned temperature for the SRC (red circles) and DRC (blue crosses, with an offset of 10 mW on the y-axis). Using the $sinc^2$ function to fit the data in Matlab, the fitting temperature bandwidth of PPKTP was $3.0K$ for both configurations. We also found that the optimum phase-matching temperature was $44.7^{\circ}C$.

Fig. 5(d) shows the single-pass conversion coefficients $\alpha$ of the PPKTP crystal with the SRC (red circles) and DRC (blue asterisks) for an increasing intra-cavity master laser power $P_1^{cav}$ (with a fixed infrared laser power). Here, the $\alpha$ was slightly higher for the SRC than for the DRC, owing to a better focusing parameter and coupling coefficient for the slaver laser for the SRC. The average values of the SRC and DRC were around $\alpha=0.05/W$, which is (to the best of our knowledge) the highest single-pass conversion efficiency with a PPKTP crystal for generating a continuous wave green laser source.
 \section{Conclusion}
In conclusion, we systematically studied the performance of SFG in both single and double resonance configurations. In section 2, we proposed a theoretical model for SFG and determined the optimized cavity parameters for both configurations. Our optimization methods can be easily expanded for other wavelengths. In the experimental section, 373 and 263 mW output power CW green lasers were achieved using both the single and double resonance configurations, and a single-pass conversion efficiency of up to $5\%/W$ was achieved for PPKTP. The generated green laser light could find wide-scale applications in quantum experiments. Using this green laser, we were able to conduct some interesting experiments, including experiments involving two color entangled photon pairs, frequency down conversion of orbital-angular-momentum-carrying laser beams, and super-resolution measurements \cite{zhou2015classical,zhou2016orbital,zhou2016orbitalPRL,zhou2017super, Liu2017Coherent}.
In the future, {\color{Revise}{this green laser should find widespread applications, for example to generate squeezed states, two color entangled photon pairs and as a tunable infrared laser in optical parameter oscillators.}}
%

\end{document}